\documentclass[12pt, a4paper]{article}
\usepackage{graphicx}

\begin{document}
\title{Sequential recruitment and combinatorial assembling of multiprotein complexes 
in transcriptional activation}
\author{Vincent Lemaire,$^{1,*}$ Chiu Fan Lee,$^{2,*}$ Jinzhi Lei,$^{3,1}$\\
Rapha\"{e}l M\'{e}tivier,$^{4}$ 
and Leon Glass$^{1}$
\\
\\
$^1$Centre for Nonlinear Dynamics, McGill University, Canada
\\
$^2$Physics Department, Clarendon Laboratory
\\
Oxford University, UK
\\
$^3$Zhou Pei-Yuan Center for Applied Mathematics
\\
 Tsinghua University, China
\\
$^4$UMR 6026, \'{E}quipe EMR, Universit\'{e} de Rennes 1, France
}

\maketitle

\begin{abstract}
In human cells, estrogenic signals induce cyclical association and dissociation of 
specific proteins with the DNA in order to activate transcription of 
estrogen-responsive genes. These oscillations can be modeled by assuming a large 
number of sequential reactions represented by linear kinetics with random kinetic 
rates. Application of the model to experimental data predicts robust binding sequences 
in which proteins associate with the DNA at several different phases of the 
oscillation. Our methods circumvent the need to derive detailed kinetic graphs, and 
are applicable to other oscillatory biological processes involving a large number of 
sequential steps.
\end{abstract}

The central dogma of molecular biology states that for a given gene,
the sequence of nucleotide bases in DNA is transcribed into messenger
RNA which in turn is translated into a specific sequence of amino
acids that constitute the protein coded by the initial gene. In higher
organisms, such as ourselves, transcriptional control is a crucial
step in the regulation of gene expression. 
This control is modulated
by the configuration of proteins around promoters (DNA regions in the
proximity of genes that carry out the integration of transcriptional
signals). Although there are a large number of theoretical models of transcriptional 
control
networks~\cite{networks} and transcriptional control of a single gene in 
prokaryotes~\cite{prokaryotes}, theoretical analysis of transcriptional control of a 
single gene in eukaryotes is much less developed~\cite{eukaryotes}.

Histones and other protein molecules associate with the DNA to form
chromatin, the constituent of the chromosomes of eukaryotes.  In order
for transcription to occur, chromatin must be unfolded from its
condensed geometry in which DNA is compactly wrapped around the
histones. Although full details are still not well understood, it is
clear that sequential chemical reactions between the histone molecules
and specialized enzymes underlie the modification of the chromatin
structure~\cite{FISCHLE2003}. For example, acetylation of the histones
leads to a more open chromatin configuration, by changing the local
electrostatic equilibrium of the molecular ensemble around where the
modification is made, enabling transcription~\cite{SUN2005}. On each histone 
protein there are a number of different
amino acids sites at which chemical reactions can occur leading to a
modification of the histone--DNA geometry. The histone code hypothesis
posits that the modifications of the histones provide a code which
governs the subsequent chemical processes leading to the remodeling of
the chromatin \cite{JENUWEIN2001}.

\begin{figure}[t]
\protect\caption{(color online). Dynamics of proteins binding at the pS2 promoter
  following administration of estradiol in \mbox{MCF-7} human breast cancer
  cells. The proportions of bound pS2
  promoters with key transcription factors and cofactors are shown as
  a function of time. Based on data from Ref.~\cite{METIVIER2003}.}
\label{data}
\includegraphics*[width=\columnwidth]{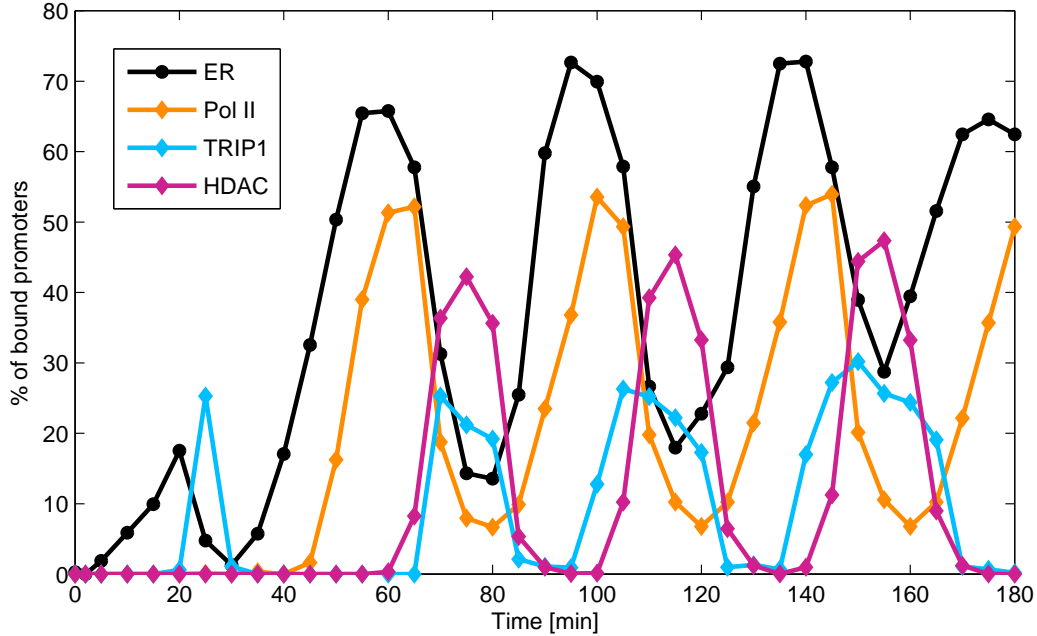}
\end{figure}

Recent experimental studies have demonstrated a
cyclic ordered sequence of reactions and alterations of
local chromatin structure in human breast cancer cells grown in tissue
culture~\cite{METIVIER2003}. The culture of approximately $2 \times
10^6$ cells is initially synchronized. The addition of a hormone,
estradiol, induces 40 min oscillations of the transcriptional
activation of the pS2 gene, which is a marker gene for estrogenic
response. Due to loss of synchronization between the cells, the observed oscillations 
slowly damp and reach constant levels after 8 hours~\cite{metivierunpub}. A possible 
source of desynchronization, in addition to stochastic fluctuations at the level of 
the promoter, could be the variability among the cells, such as ATP levels or cell 
size. These oscillations are monitored by measuring the temporal
association of specific proteins with the DNA measured at time intervals of as short 
as 5
minutes over a 3 hour period~\cite{chipfootnote}. In Fig.~\ref{data}, we show the 
association profiles of four
key proteins involved in the transcriptional activation of gene
pS2. Estrogen receptor (ER) binds estradiol and initiates the transcription process. 
RNA
Polymerase~I\hspace{-.06em}I, (Pol~I\hspace{-.06em}I) is a protein
complex responsible for the transcription of genes. TRIP1 and HDAC are
two different proteins that are involved in the 
clearance of the promoter after each
transcription cycle. In view of the complexity of the
sequence, and the tiny numbers of molecules involved in the binding in
each cell, it is currently impossible to derive detailed kinetic data
about the rate constants of the individual reactions. Moreover, since
it is not clear whether or not the cells are coupled to each other,
the mechanism of the synchronization of the oscillation poses a
challenge for theoretical interpretation. In this Letter, we propose
a simple model for the oscillation based on a large number of
sequential chemical reactions and transformations of the chromatin. 
Based on the analysis of the model, we are able to predict specific
timings of the association of the protein complexes with the chromatin
that reproduces the observed dynamics in Fig.~\ref{data}.

\begin{figure}[t]
\caption{(color online). A schematic diagram of the model by sequential recruitment of
  protein molecules to the chromatin.  $x_1$ represents the chromatin
  at the pS2 promoter. The $x_i, \,\, 2
  \le i \le m$ represent the protein complexes that form successively on
  the promoter, at rate $a_i$, leading to the activation of 
transcription.}\label{sequence}
\includegraphics*[width=\columnwidth]{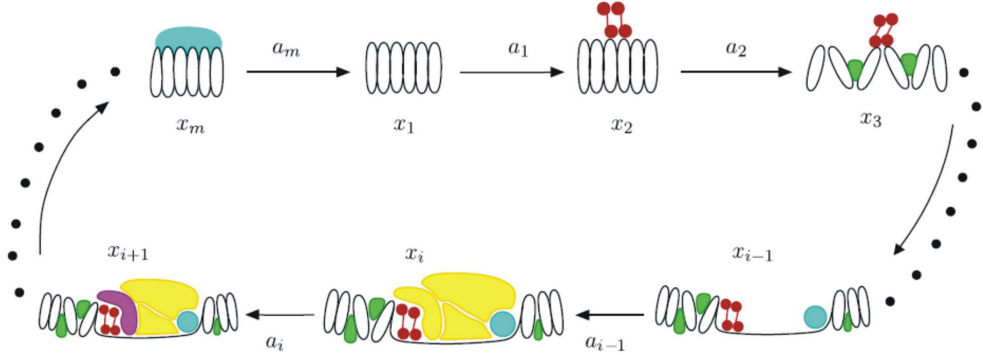}
\end{figure}

We assume that there is a network of proteins interacting together
and, sequentially, with the chromatin. Each reaction induces a
modification of the substrate complex that in turn enables the next
step in the sequence so that the reactions are assumed to be
irreversible. 
Further, we assume that the various transcription factors and
cofactors involved in the reactions are present in sufficiently high
concentration that the reaction rates $a_i$ are constant over
time. This model is schematized in~Fig.~\ref{sequence}.
A key parameter in our model is the number $m$ of sequential steps
in the cycle.  Many transcription complexes contain more than 50
proteins, which may be partially or completely assembled on the
promoter~\cite{refspack}.  Of the order of 100
histone (or other proteins) modifications have been identified during
transcriptional activation~\cite{LEE2005}. These sequential histone
modifications are associated with the histone code for transcriptional
activation. Based on these considerations, we estimate that $m$ is at
least 200.  From the data in Fig.~\ref{data}, the period of the
oscillation $T_0$ is about 40 minutes. If all reaction rates are
assumed identical (i.e., $a_i=a$), and choosing $m=200$, then we have
$a=m/T_0=5 {\rm min}^{-1}$.

\begin{figure}[t]
\protect\caption{(color online). A-C: Stochastic simulations for 1, 10, or 100 cells.
D:~Recruitment curves calculated from~Eq.~(\ref{protein}). These
computations are based on the
binding sequences given in Fig.~\ref{seq} (see below). 
We assume $a_i=5 {\rm min}^{-1}$ and
$m=200$.  The color (or grayscale) code is the same as in Fig.~\ref{data}.}
\label{sto}
\includegraphics*[bb=108 236 502 555,width=\columnwidth]{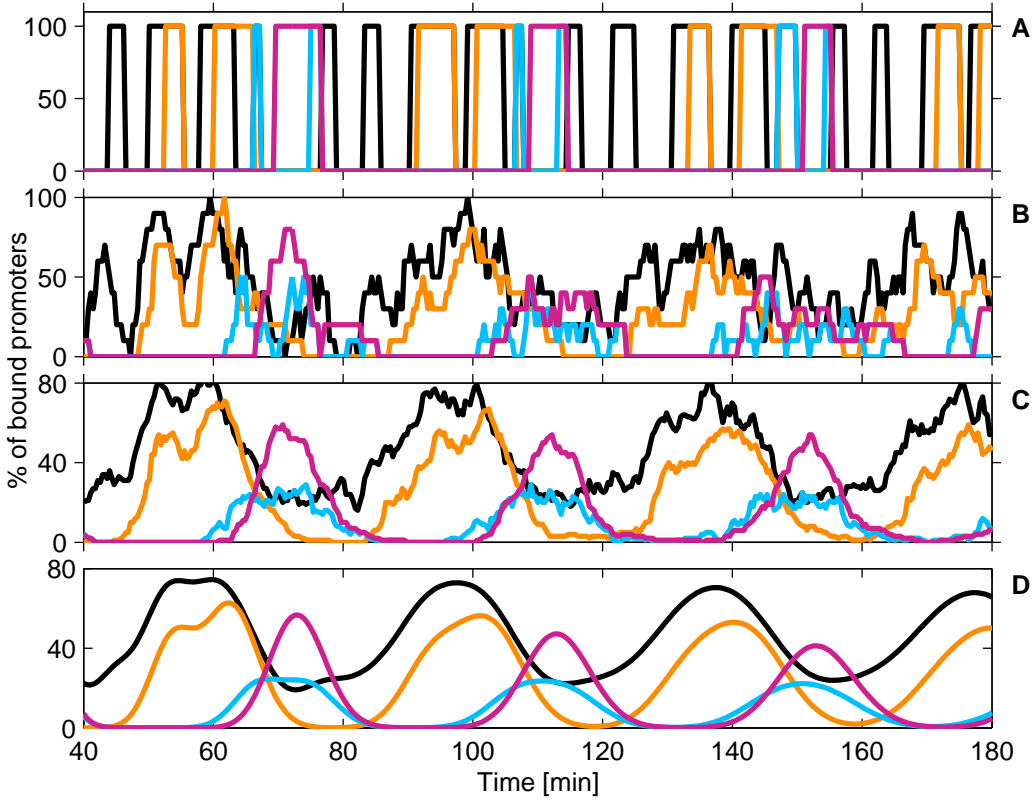}
\end{figure}

Since, generally, there are only two copies of each gene in each cell,
we first consider the dynamics using a stochastic model in which the 
probability of a given reaction per unit time is equal 
to the product of the rate constant for that reaction and the number 
of potential reactants present.  The time steps between reactions obey a Markov 
process. The
results of carrying out the simulation using the Gillespie
algorithm~\cite{GILLESPIE1977}, for 1, 10, or 100 cells are shown in
Fig.~\ref{sto}A-C. 

The chemical scheme presented in Fig.~\ref{sequence} can be
interpreted as a sequential Poisson process in which the duration $t$
before the next reaction takes place follows the distribution
$p(t)=a\mathrm{e}^{-at}$. Then for any individual cell after synchronization,
the $k$-th cycle has a mean starting time of $\frac{km}{a}$ with a
variance $\frac{km}{a^2}$. Taking $a=m/T_0$, the starting time of the
$k$-th cycle has a variance of $\frac{kT_0^2}{m}$. This suggests the
natural desynchronization of the system as the variance increases
linearly with $k$. As $m \rightarrow \infty$, the variance vanishes
and the system behaves as a delay differential system as discussed in
Ref.~\cite{MACDONALD1989}. 
Thus, just as in the data in Fig.~\ref{data}, the stochastic system displays 
oscillations for
several cycles provided $m$ is large enough.

In the limit of a large number of cells, the stochastic dynamics can
be approximated by the linear differential equations
\begin{eqnarray*}
\dot{x}_1(t) &=& a_mx_m(t)-a_1x_1(t)\label{model_1}\ ,\\
\dot{x}_i(t) &=& a_{i-1}x_{i-1}(t)-a_ix_i(t)\label{model_2}\ , \ \ 2\leq i \leq m.
\end{eqnarray*}

In the case $a_i=a$, the eigenvalues $\lambda_k$ of the Jacobian matrix
are $\lambda_k=a\ \bigl(\mathrm{e}^{\frac{2\pi k i}{m}} - 1\bigr),
\,1\leq k \leq m$. We can rewrite $\lambda_k$ as $\alpha_k + i\beta_k$, where
\begin{equation}
\alpha_k = a\ \bigl(\cos\theta_k - 1\bigr), \,\,\beta_k = a\
\sin\theta_k,\,\, 1\leq k \leq m\ .
\label{eigenvalues2}
\end{equation}
Here $\theta_k=2k\pi/m$. Since the real parts of the eigenvalues are
non-positive, Eqs.~(\ref{model_1})-(\ref{model_2}) do not show
sustained oscillations~\cite{HEARON1953}. 

Assuming initial conditions, for the $[x_i]$, of 
$[1,0,0,\ldots,0]$, we can compute the solution for all
variables:
\begin{equation}
x_i(t) = \frac{1}{m} \left[  1 +(-1)^i e^{-2 a t} + 2\sum_{k = 1}^{m/2 - 1} 
e^{\alpha_k t} \cos [\beta_k t - (i-1) \theta_k] \right],\label{exact2}
\end{equation}
in the case when $m$ is even (a similar expression holds when $m$ is odd). The higher 
frequency terms decrease
rapidly so that for $m=200$, only the first 6 or 7 terms give a
significant contribution after the first period.  The leading term 
$\mathrm{e}^{\alpha_1
t}\cos\beta_1 t$ sets the period. From a Taylor expansion of this
result, we find that the envelope of the leading term decays as
$\mathrm{e}^{-2\pi^2t/mT_0}$, where we set $a=m/T_0$. This result is consistent with 
the
finding that the oscillations are more persistent as the number of
steps of the reaction increases. 

We now consider the effect of relaxing several of the unrealistic
assumptions in the model. If all reactions are reversible, with all
forward rate coefficients equal to $a$ and all backward reaction
coefficients equal to $b$, the real and imaginary
parts of the eigenvalues are $\alpha_k =
(a+b)\bigl(\cos\frac{2k\pi}{m} - 1\bigr)$ and $\beta_k = (a-b)
\sin\frac{2k\pi}{m}$.  This leads to an increased damping, and an increase in the 
oscillatory period that scales as $b/a$ for small $b$
in comparison to $a$. Consequently, the main results presented below also hold if the 
reactions are reversible. A more general discussion on the effects of
reversible reactions in chains of linear reaction kinetics is given
in Ref.~\cite{SUMMERS1988}.

Since in the biological system, the reaction rates are not identical, we
now assume that the forward rates are distributed randomly with probability 
distribution $Q(a_i)$.  Realizing that the waiting time for each
reaction to occur is independent and identically distributed, the
$k$-th cycle's starting time has mean and variance, $km \int
\frac{Q(x)}{x}dx$ and $km \int\frac{Q(x)}{x^2}dx$, respectively. If, in
particular, $Q(a_i)$ is the uniform distribution over the interval
$[a(1-d),a(1+d)]$, with $0\leq d\leq 1$, the
$k$-th cycle's starting time has mean:
\begin{equation}
\frac{km}{2ad} \int_{a(1-d)}^{a(1+d)} \frac{dx}{x} = \frac{km}{2ad} \ln 
\frac{1+d}{1-d}\ ,
\end{equation}
and variance:
\begin{equation}
\frac{km}{2ad} \int_{a(1-d)}^{a(1+d)} \frac{dx}{x^2} = \frac{km}{a^2(1-d^2)}\ .
\end{equation}
Then the mean and variance of the k-th period are, respectively,
$\langle T_k\rangle=\langle T\rangle=\frac{m}{2ad} \ln
\frac{1+d}{1-d}$ (independent of $k$) and
$\sigma_{T_k}^2=\frac{(2k-1)m}{a^2(1-d^2)}$. Thus, on the curve
$\langle T\rangle=T_0$, we have the variance
$\sigma_{T_k}^2=\frac{4(2k-1){T_0}^2d^2}{m\ln[(1+d)/(1-d)]^2}$,
which, again, vanishes as $m\to\infty$. 

The damping of the solutions of
\mbox{Eqs.~(\ref{model_1})-(\ref{model_2})} is controlled by the least
negative $\alpha_k$'s. Let $a_i = a (1 + \epsilon \omega_i)$ and
$\epsilon = \sigma_a/a$, where $\sigma_a$ is the standard deviation of the
$a_i$'s.  Assuming that $a_i$ is uniformly distributed in the interval
$[a(1-d),a(1+d)]$, we have $\langle a_i\rangle = a$
and $\sigma_a=ad/\sqrt{3}$, and then, $\langle \omega_i\rangle = 0$
and $\langle \omega_i^2\rangle = 1$. Solving the characteristic
equation for successive orders in $\epsilon$, we find that, 
to fourth order in $\epsilon$,  
$\langle \alpha_k\rangle = a(\cos \theta_k -1)(1+o(\epsilon^4))\ ,$ 
which is consistent with numerical results that show negligible
dependence of $\langle \alpha_k\rangle$ on $d$. 
However, the variance of the $\alpha_k$'s is 
$\sigma_{\alpha_k}^2 = \frac{\langle \alpha_k\rangle^2}{m}\epsilon^2+o(\epsilon^4)\ .$
This shows that the properties of low damping and synchronization of
the oscillation, observed when the $a_i$'s are identical, are conserved in the limit 
of large $m$ when the rate constants are different. 

We now wish to fit the model to the experimentally observed binding
profiles of the proteins in Fig.~\ref{data}.  Each protein is a
component of several different $x_i(t)$ complexes, but we do not know
a priori which ones.  We call $P_j(t)$ the percentage 
of pS2 promoters bound to one or more molecules of protein
$j$. Then we have 
\begin{equation}
\label{protein}
P_j(t) =\sum_{i=1}^{m} c_{i,j} x_i (t), 
\end{equation} where $c_{i,j}$ is either 0 or 1.  
$P_j(t)$ is the quantity measured in ChIP experiments shown in Fig.~\ref{data}. 
For each protein $j$, the {\it binding sequence} 
$\{c_{i,j}\}$ is determined by doing a least squares
minimization of the data to the model. Because the first cycle is produced by a
different sequence of chemical steps than the subsequent
ones~\cite{METIVIER2003}, we consider only the time points such that
$T_0\le t \le 3T_0$. The minimization procedure is done in 2
steps: (i) we apply the Nelder-Mead method to minimize the quadratic
error, with the constraint that $0\leq c_{i,j}\leq
1$~\cite{NELDER1965}; (ii) we use the values of $c_{i,j}$ obtained in the
first step as initial conditions to a method that uses Lagrange
multipliers, minimizing again the mean square error. The latter step enables us to 
generate binary $c_{i,j}$'s. 
\begin{figure}[t]
\caption{(color online). Binding sequences for the proteins in Fig.~\ref{data}. For 
each protein, the colored regions indicate the indices $i$ of the complexes $x_i$ of 
which the protein is a component.}
\label{seq}
\includegraphics*[width=\columnwidth]{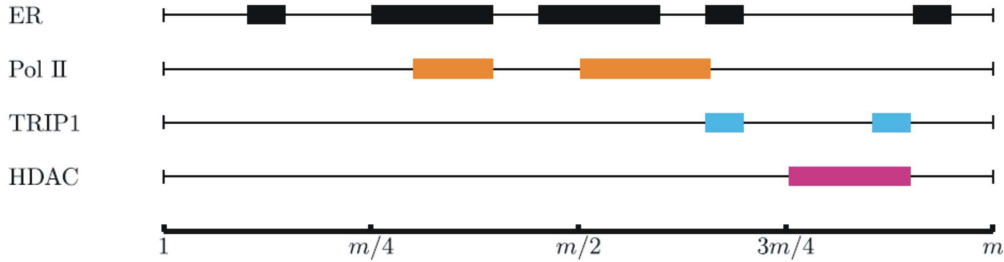}
\end{figure}
The result of that procedure is
shown in the Fig.~\ref{seq}, where the colored regions indicate the
regions where $c_{i,j}=1$. The $P_j(t)$, for each protein in Fig.~\ref{data}, are 
plotted in Fig.~\ref{sto}D.
To test the robustness of these fitted sequences, we have carried out
a number of numerical studies in which we performed fits of the model
to the data relaxing several of our assumptions. In particular, we
have tested for values of $m=100$, 200, 300 and 400; addition of $\pm
2$\% of noise to the data points (corresponding to the error reported
in Ref.~\cite{METIVIER2003}); changes in the vertical scaling of the data
(up to 1.4); and selection of random reaction rates $a_i$ (provided that the
period, for the selection of $\{a_i\}$, is close to $T_0$, and the solutions not too 
damped). 
Although there can be slight changes in the values of $i$ where
$c_{i,j}=1$, or, in some circumstances, a change in the number of blocks in which
$c_{i,j}=1$, the main pattern of the $c_{i,j}$'s in
Fig.~\ref{seq} remain unchanged. 

The results in Fig.~\ref{seq}, in which the precise patterns of
association of each protein are obtained, represent the main predictions of the 
current work. Although one might have 
anticipated that there would be a single recruitment block for each
protein, the recruitment patterns for proteins considered here 
may actually occur in two or more blocks. 
Two experimental methods can be used currently to determine the dynamics of 
transcription: ChIP assays and fluorescence microscopy-based assays, but these methods 
seem to produce conflicting results~\cite{HAGER2004}. Our theoretical predictions must 
be viewed in perspective of these two experimental methods. ChIP assays determine the 
binding of promoters with specific proteins, but not at the level of one promoter in a 
single cell. In contrast, fluorescence microscopy assays determine the mobility of 
proteins around individual promoters, but does not measure binding of individual 
proteins to one promoter. Using ChIP data, our model enables us to predict successive 
rounds of protein binding and unbinding.  The protein mobility indicated by these 
multiple binding events, corroborates the observations from fluorescence microscopy 
assays, reconciling the observations from the two experimental methods. Due to lack of 
precision at the scale of the isolated promoter, the experimental verification of our 
findings is currently impossible. However, these results can be correlated with what 
is known about the biological system. For example, five or six principal complexes, 
with well-defined functions, are successively formed on the promoter of 
pS2~\cite{METIVIER2003}. It has been conjectured that the assembling of these 
complexes is orchestrated by ER at different timings of the 
cycle~\cite{METIVIER2003,LEE2005}. Our finding of five different binding times of ER 
matches perfectly that conjecture.

To summarize, we have proposed a simple model for the oscillation observed in protein 
association and dissociation during transcriptional activation in human cells. We have 
shown that the model produces oscillations with minimal damping for large values of 
$m$. Further, these properties are conserved when the reaction rates are selected 
randomly. The current work demonstrates that realistic network architecture models may 
not be needed in order to unravel the
mechanisms of complex reaction sequences at the subcellular level. Our approach relies 
rather on
the finding that synchronous dynamics of protein assembly emerge as a consequence of 
the large number of intermediate reactions. Our methods should be useful to other 
systems in which many sequential steps take place but the detailed kinetics are not 
known. Fitting the model to the data in Fig.~\ref{data} 
resulted in predicted sequences at a time resolution not possible
experimentally and, as such, may be invaluable for experimental 
design and for interpretation of the mechanisms underlying
transcriptional 
activation. 

\vskip.5in
This research has been partially supported by NSERC. We thank Dr. John
White and Luz Tavera Mendoza, McGill University for helpful
conversations. CFL thanks University College (Oxford) for financial
support. 
\\
\\
\small{$^*$The first two authors have contributed equally to the present work.\\
Corresponding author: Vincent Lemaire (lemaire@cnd.mcgill.ca)}


\begin{thebibliography}{19}
\bibitem{chipfootnote} 
The association is monitored using chromatin immunoprecipitation assay (ChIP) and 
quantitative polymerase chain reaction (PCR). This allows the isolation and 
identification of proteins that are bound to a specific region of the DNA.
\end{thebibliography}
\end{document}